\documentstyle[dina4, epsf]{article}
\begin{document}

%
\newcommand{\xsize}{11cm}
\newcommand{\bitxsize}{15.1cm}
%
\newcommand{\aff}[1]{${}^{#1}$}
\newcommand{\org}[2]{~\\\footnotesize \aff{#1} #2}
\newcommand{\organizations}[1]{\date{\footnotesize #1~\\[0.5cm]\large\today}}
\newcommand{\keywords}[1]{\noindent{\em Keywords:} {\bf #1}}
\newcommand{\ZPR}{Center for Parallel Computing, Cologne University,
D-50923 K\"oln, Germany}
\newcommand{\TSASA}{Los Alamos National Laboratory, TSA-DO/SA MS M997, Los Alamos NM 87545, USA}
\newcommand{\SFI}{Santa Fe Institute, 1399 Hyde Park Rd, Santa Fe NM
87501, U.S.A.}
\newcommand{\tempname}{temp}
\newcommand{\tempb}{temp}
\newenvironment{tab}[3]{\renewcommand{\tempname}{#2}\renewcommand{\tempb}{#3}%
\begin{table}\begin{center}%
\begin{tabular}{#1}}%
{\end{tabular} \caption{\label{\tempb} \em\tempname}\end{center}\end{table}}
\renewcommand{\bottomfraction}{1.0}
\renewcommand{\topfraction}{1.0}
\renewcommand{\textfraction}{0.0}
\setlength{\floatsep}{0.5ex plus0.25ex minus0.25ex}
\setlength{\textfloatsep}{0.5ex plus0.25ex minus0.25ex}
\setlength{\intextsep}{0.5ex plus0.25ex minus0.25ex}
\newcommand{\eps}[3]{\begin{figure}\begin{center}\leavevmode\epsfbox{#1}\caption[{\em
#2}]{\label{#3}{\em #2}}\end{center}\end{figure}}
%
%

\thispagestyle{empty}
\title{Experiences with a simplified microsimulation\\
for the Dallas/Fort Worth area}
\author{M.\ Rickert\aff{a,b} and K.\ Nagel\aff{b,c}}
\organizations%
{
\org{a}{\ZPR\\mr@zpr.uni-koeln.de}
\org{b}{\TSASA\\rickert@lanl.gov kai@lanl.gov}
\org{c}{\SFI\\kai@santafe.edu}
}
\maketitle
\begin{abstract}
We describe a simple framework for microsimulation of city traffic. A
medium sized excerpt of Dallas was used to examine different levels of
simulation fidelity of a cellular automaton method for the traffic flow
simulation and a simple intersection model. We point out problems
arising with the granular structure of the underlying rules of motion.

\vspace{0.5cm}
\end{abstract}
\newpage
\section{Introduction}
In recent time, traffic demand in metropolitan areas has largely
exceeded capacity. Typical remedies such as the expansion of the road
way system or road improvement do not work well anymore for various
reasons: On the one hand communities cannot afford the increased costs
of road construction, on the other hand there are political and
environmental objections.

Therefore, computer simulation as a means of evaluating, planning, and
controlling large traffic systems has gained considerable
importance. The classical approach of traffic research consists of
four steps: traffic generation, route planning, static assignment of
routes onto the network, and evaluation. In recent years, more and
more work has been invested to consider the dynamic aspects of these
components. Especially the microsimulation part has taken great
advantage of models developed in classical research areas such as
physics, mathematics and computer science.

The failure of classical simulations to describe (and possibly
predict) traffic was in part due to insufficient computational
speed. One had the choice of simulating a rather large area (e.g.\ the
metropolitan area of Los Angeles) at a low fidelity (vehicle densities
instead of individual vehicles) {\em or\/} simulating a small excerpt
(e.g. a suburb) at a high resolution. Only in recent years, it has
become possible to use high-performance computers in conjunction with
computationally efficient microsimulation models to simulate large
areas at high resolution. Despite this advantageous development,
execution speed still remains an issue. Especially since the traffic
simulation consists of several stages, some of which have to be
iterated several times to produce valid and consistent results. In
this paper we would like to concentrate on the microsimulation
portion.

In the next section, we present a simple model capable of executing a route
set provided by a router. The functionality of the two major components,
namely intersections and street segments, are outlined. The enforcement of
speed limits and the usage of traffic lights are used to define four
different levels of fidelity. We continue in section\ \ref{results} by
showing first results of the simulation obtained in conjunction with a
TRANSIMS \cite{NaBarrRi:1} case-study. Section \ref{performance} outlines
the computational performance of the model as well as some aspects of its
implementation on a parallel computer. We conclude by giving a short
summary and some hints at future work.

\section{Model}
\label{model}
We use a simple representation of the underlying street network by
transferring its basic elements (e.g. intersections and street
segments) into their simulation counterparts, which we call nodes and
edges. Each intersection is associated with a node. In contrast to
their real-world counterparts, the simulation nodes do not have a
micro-structure (e.g. shape, elaborate traffic light phases), but are
reduced to very basic characteristics. 

Each direction of street segment is represented by a directed edge
connecting the respective nodes and pointing into the direction of
traffic flow. This convention covers both one-way (e.g. exit/access
ramps of freeways) and bi-directional street segments. Internally the
traffic on each edge is simulated with a modified version of the
cellular automaton of Ref.~\cite{Nagel.92.Schreck}.

Next we will describe the two major components of our simulation.

\subsection{Street Segments}
The directed connection (edge) between two nodes is represented as
a grid equivalent to the model by Nagel/Schreckenberg
\cite{Nagel.92.Schreck} and its two-lane extension
\cite{Rickert.96.etc.twolane}. The characteristics length\footnote{The
length of a street segment is either explicitly given or derived from
the Euclidean distance of the two nodes.}, speed-limit, and number of
lanes are used to adapt the cellular automaton (CA) model. The size of
the grid is computed by using the grid-site length of 7.5~[meter] as a
unit.  

It is important to note that the characteristics mentioned so far are
{\em constant} for the whole segment. Typical details like additional
turning lanes in front of intersections are modeled by inserting
additional nodes to split a given segment and assigning different
parameters to the various parts.

\subsubsection*{Cellular Automaton Model}

In case of only one lane the model of motion is equivalent to the original
Nagel/Schreckenberg traffic CA~\cite{Nagel.92.Schreck} with an optional
speed-limit $v_{sl}<v_{max}$.  We would like to outline this model for the
convenience of the reader.

The system consists of a one dimensional grid of $L$ sites with
periodic boundary conditions. A site can either be empty, or occupied
by a vehicle with an integer velocity zero to $v_{max}$. The velocity
is equivalent to the number of sites that a vehicle advances in one
update --- provided that there are no obstacles ahead. Vehicles move
only in one direction.  The index $i$ denotes the index of a vehicle,
$x(i)$ its position, $v(i)$ its current velocity, $v_{sl}(i)$ (``sl''
for ``speed limit'') its maximum speed imposed by a segment-specific
speed-limit\footnote{Note that in the original model all vehicles had
the same maximum velocity $v_{max}$} $pred(i)$ the index of the
preceding vehicle,\footnote{%
{\em A precedes B} in this context means
that {\em A} is followed by {\em B}.
} $gap(i):=x(pred(i))-x(i)-1$ the width of the gap to the
predecessor. At the beginning of each time-step rules $S1\ldots S3$
(see below) are applied to all vehicles simultaneously (parallel
update, in contrast to sequential updates which yield slightly
different results).  Then the vehicles are advanced according to their
new velocities.
\begin{itemize}

\item 
{\bf IF} $v(i) < v_{sl}(i)$ {\bf THEN} $v(i) := v(i)+1\quad$ ({\bf
S1})

\item 
{\bf IF} $v(i)>gap(i)$ {\bf THEN} $v(i) := gap(i)\quad$ ({\bf S2})

\item 
{\bf IF} $v(i)>0$ {\bf AND} $rand<p_d$ {\bf THEN} $v(i) := v(i)-1\quad$ ({\bf S3})

\end{itemize}
{\bf S1} represents a constant acceleration until the vehicle has
reached its maximum velocity $v_{sl}$. {\bf S2} ensures that vehicles
having predecessors in their way slow down in order not to run into
them. In {\bf S3} a random generator is used to decelerate a vehicle
with a certain probability modelling erratic driver behavior. The
free--flow average velocity is $v_{sl}-p_d$ (for $p_d\ne 1$).

In case of two lanes the model corresponds to the one described in
\cite{Rickert.96.etc.twolane}. For more than two lanes we extend 
the model using the same lane-changing rules. Collision of vehicles
changing to the same neighboring site are prevented by a right-before-left
lane priority (see also \cite{RiWa:1}).

Recently, the strictness of the CA model has been relaxed by
maintaining only time as a discrete value, while using continuous
values for location and velocity \cite{Wagner.96.mlane}. Modifications
to the multi-lane rule-set yielded promising similarity to the
lane-changing behavior found on German Autobahn through-ways
\cite{NaWaWo.97,WaNaWo.97,Wagner.95.julich}.
\subsubsection*{Speed Limit}
In contrast to the original CA model which assumed a maximum limit of
approximately 120 km/h (freeway traffic), the speed limit within a
city is usually lower. In order to match the speed-limits of the
Dallas study area, for each segment we introduced a CA speed limit
\begin{displaymath}
v_{sl} = \lfloor v_{sl}^{real} / l_{site} + 0.5 + p_d\rfloor
\end{displaymath}
where $v_{sl}$ is the CA speed limit given in sites per time-step,
$v_{sl}^{real}$ the real speed limit given in meters per second,
$l_{site}$ the CA site length given in meters, and
$p_{d}$ the deceleration probability of the CA rule set. Moreover,
$v_{sl}$ is forced to be in $[1\ldots v_{max}]$.

\subsection{Intersections}

\epsfxsize=\xsize
\eps{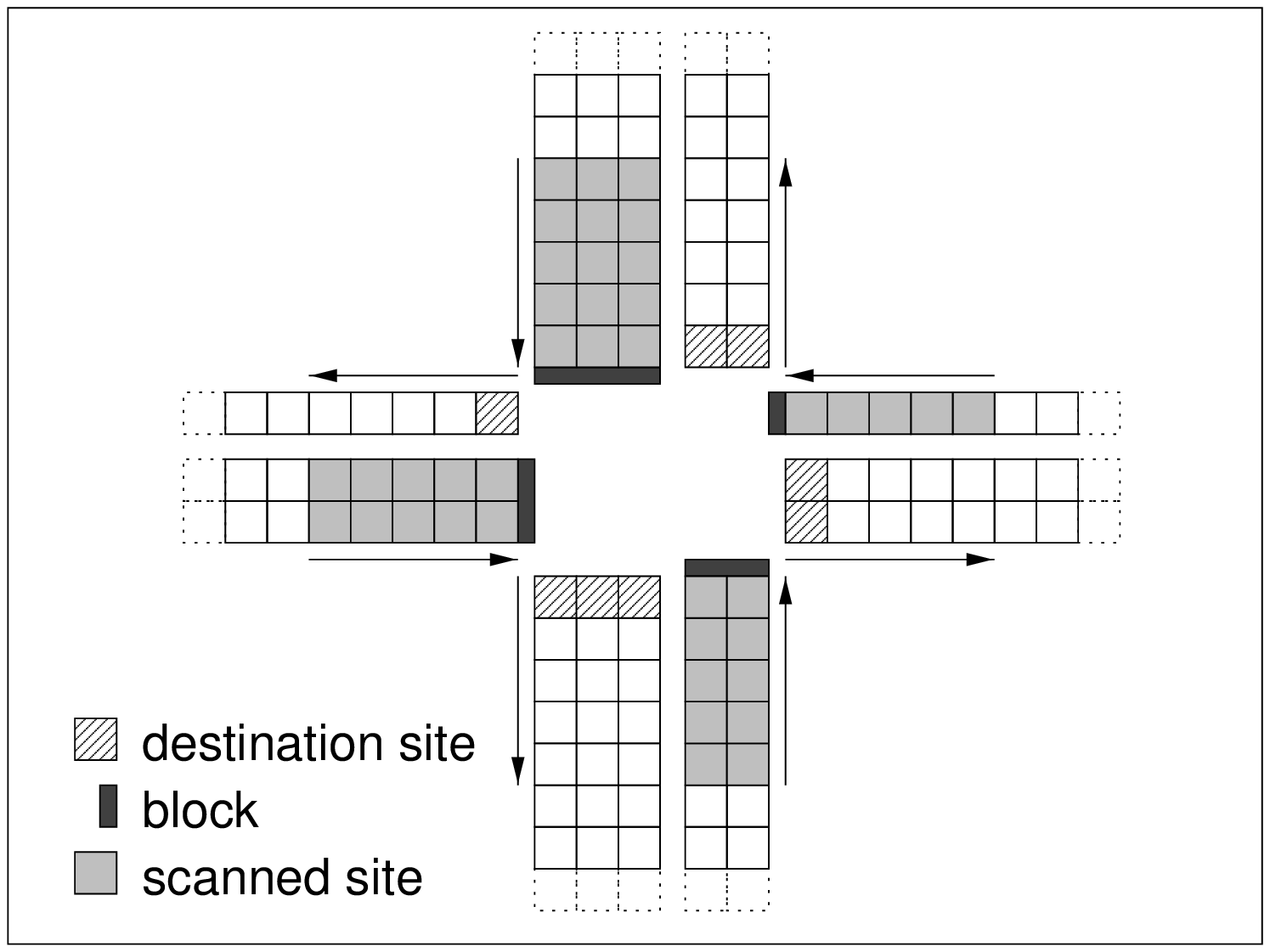}{Geometry of an intersection}{fig_intersection_geometry}

Intersections are modeled in a very simple fashion. The geometry is
shown in figure \ref{fig_intersection_geometry}. Although the figure
gives the impression of an intersection having a spatial 'extent', the
simulation actually treats it as having no size at all.  A vehicle is
never {\em inside} the intersection, but either on the incoming lane
or on one of the outgoing lanes. Therefore, the intersection does not
have an explicit capacity unless incorporated into the network
(e.g. transfer lanes at freeway-junctions \cite{RiWa:1}). An
intersection of degree two (usually a node used to mimic the geometric
trajectory of a street) is treated in a special way by connecting the
incoming segments without any interruption visible to the vehicles.

All incoming lanes of each incoming segment are equivalent.
At the very end of each incoming lane $v_{max}$ sites are scanned for
vehicles {\em before} the usual rules of motion are applied. During
each time-step, at most one vehicle per incoming lane of the source segment
can be moved to one of the insertion sites of the destination
segment. If possible the same lane is used on the destination segment. If the destination
segment has fewer outgoing lanes than the incoming segment has
incoming lanes the vehicle is inserted into the leftmost lane. If that
site is occupied, the next neighboring site off to the right is checked until a
vacant site is found or the right-most lane is reached. Note that the
scanning of the incoming lanes is always done beginning at the site nearest
to the intersection. That way, the order of the vehicles with
respect to each other is not changed. In order to ensure unbiased processing
of all incoming lanes, the scanning is done in a robin round fashion 
with respect to consecutive time-steps.

For a vehicle approaching the intersection there are two alternatives:
either it is absorbed from the scanning area and inserted into the
destination segment or it proceeds according to its CA rules of
motion. Due to block at the end of the lane (or earlier due to other
preceding travelers) it will eventually stop. This behavior is
important to model the spill backs in real world traffic. The queues
are resolved as one would expect: As soon as the situation on the
destination lane(s) improves, vehicles are removed one by one starting
at the site nearest to the block. Note that a vehicle may be blocked
by other vehicles (having a different destination) although its
destination segment is vacant.

\subsubsection*{Approach and turning behavior}
In contrast to other traffic simulations with resolution at
city-street level, we do not model a special behavior for vehicles
while approaching or transversing intersections. In the implementation
described in this paper, all incoming lanes are equivalent. This was
done for two reasons. First, modeling detailed approach and turning behavior
requires extensive geometric information which is often not available
or not consistent. Second, the current rules of motion show a quickly
decreasing lane-changing probability as soon as the density exceeds a
certain threshold. This is mainly due to a strict 'look-back' rule
which checks for following traffic on the neighboring lane. In
contrast to freeway conditions, where this rule maintains the desired
traffic jam waves (see \cite{Rickert.96.etc.twolane}), in this
context, it would prevent proper lane-changing. This again would
result in vehicles queuing up, since they could not change to their
respective turning lanes \cite{Barrett.personal}.

\subsubsection*{Traffic Lights}
Traffic lights are modeled by activating the scanning mechanism for
the duration of the green phase and deactivating it for the length of
the non-green-phase (which includes both the red phase and transition
phases). Since there is only one phase per incoming segment, any
direction-specific phasing information is averaged over all directions
weighted by the number of active lanes into the respective direction.

Note that due to this averaging, the complete phase cycles
$T_{rg}=T_g+T_r$ (green phase + red phase) of the incoming segments
may differ from each other, resulting in a continuous phase shift.
This is different from real world traffic light installations where
the starting time is usually defined by taking a multiple of $T_{rg}$
and adding a relative offset.

\subsubsection*{Interferences}
Two types of interference can occur at an intersection: (a)
vehicles that have to obey right of way must wait for gaps in the
crossing or oncoming traffic stream, and (b) vehicles that have right
of way are obstructed by others blocking the intersection. In the
simplest version of our intersection of neither of these interferences
is handled. However, it is simple to force a reduced throughput
through the intersection by examining the overall occupancy of the
$v_{max}$ last sites of all outgoing segments and introducing an
additional transfer probability. This probability would be one if all
sites in the examined area are vacant, zero if all occupied with a
functional (possibly linear) transition between the two extremes.

\subsection{Plan Following}
For each simulation run all plans can be regarded as static. For the time
being, we do not perform any on-line re-routing. A plan-set is generated
from an activity set consisting of a source plus a departure time on the
one hand and destination on the other hand.  The plan-set used for the study
presented here (also referred to as {\em plan-set 11}) is a very preliminary
plan-set which was generated in the
course of the Dallas/Fort Worth case study of TRANSIMS.  Since this paper
concentrates on microsimulation aspects, describing the method how plan-sets
were obtained goes beyond the scope of this paper.  Work on plan-sets in the
context of the same case study can be found in
Refs.~\cite{Marathe:etc:DFW-planner, Nagel:feedback}.

The plan-sets in TRANSIMS are sequences of links plus estimates when
these links will be reached.  From this, one can calculate when every
driver {\em expects\/} to be on a certain link.  In consequence, one
can calculate the complete time dependent {\em demand\/} structure,
which would correspond to the traffic which would happen if everybody
could act according to her expectations.

This demand structure is best displayed in terms of the ratio of flow
demand to capacity, also called V/C ratio.\footnote{%
V stands for volume, C for capacity.  One has to keep in mind that
V refers to volume {\em demand}.
} If the V/C ratio is exactly one, the number of vehicles which are
planning to use a certain link is equal to the maximum number of
vehicles the link can let through.  If the V/C ratio is larger than
one, then the number of vehicles in excess of the capacity will not be
able to get through and will be queued up at the entrance of the link.
This means that these vehicles will not be at other links downstream
when they expected, so that the demand structure downstream of a
congested link (i.e.\ V/C $> 1$) becomes wrong.  From this argument it
becomes immediately clear that one has to run the microsimulation to
obtain dynamically correct traffic patterns.

\epsfxsize=\bitxsize
\eps{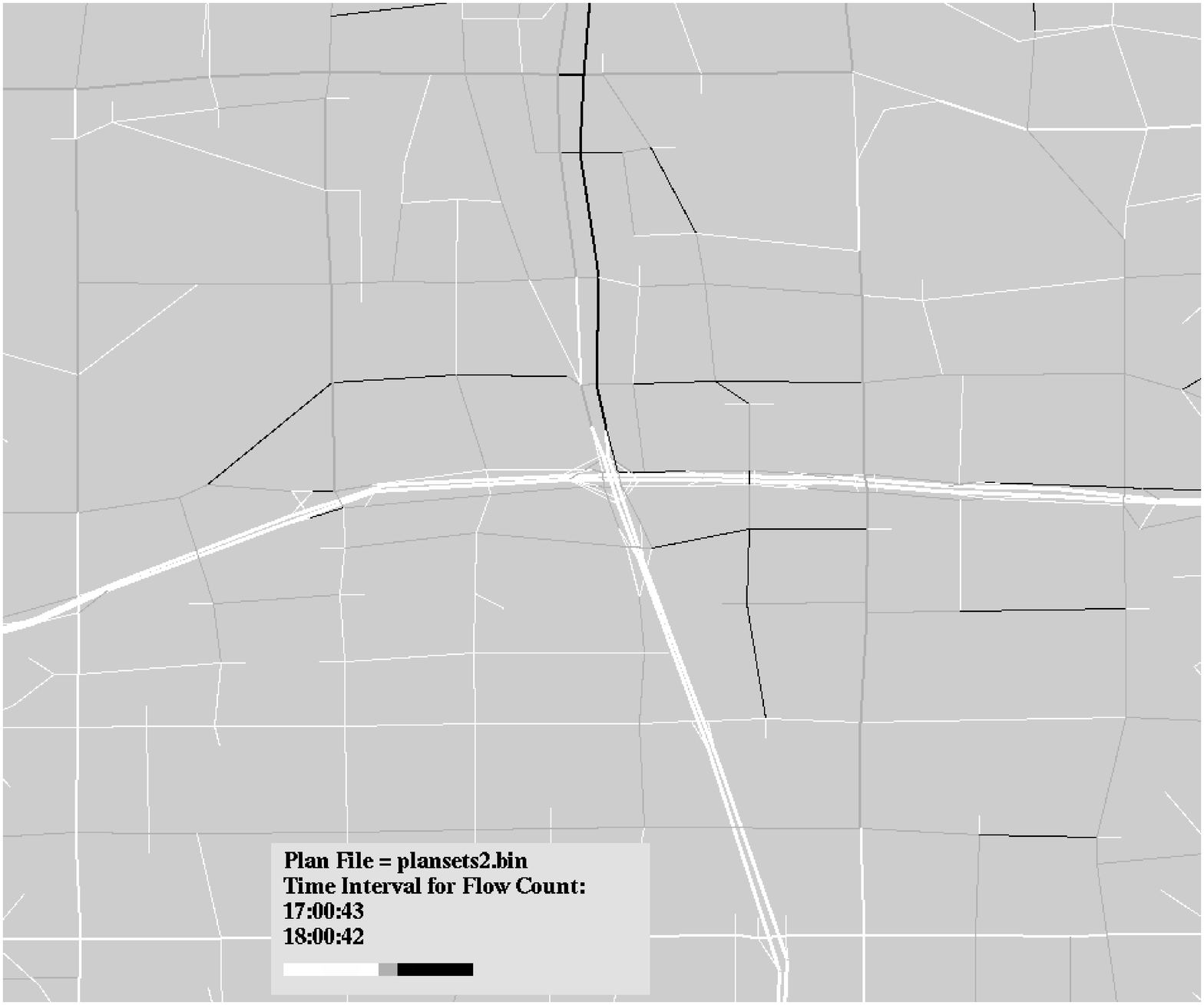}{V/C-ratios.  White links mean that demand is lower than
capacity ($V/C < 1$); for gray links demand is between capacity and two
times capacity ($1 \le V/C < 2$); for black links demand is larger than two
times capacity ($2 \le V/C$)}{fp-2}

Fig.~\ref{fp-2} shows a graphical display of the demand structure of a
plan-set which is similar to the one used for the simulation runs in
this paper.

All plan-sets are computed for the whole Dallas/Forth-Worth area which
means that all routes have to be restricted to the study area if only
that portion of the map is simulated. The truncation of the plans is
done in a straight-forward way: any route that contains at least one
segment within the study-area will be part of the restricted plan
set. Its departure will be delayed by the amount of time that the
vehicle would spent outside the study-area before it reaches the first
segment within the simulation area. For all edges transversed up to
entry, we use the cruising velocities assumed by the planner.

After the start of the simulation, route-plans are executed as
follows: (a) At the time-step given by the departure-time, a vehicle
is created at the departure node (source) of the route. (b) The
vehicle is inserted into a queue associated with the source. (c) Each
time-step the queue is scanned for pending vehicles. If possible, the
vehicle is removed from the queue and inserted into the first segment,
from where it starts following its route-plan. (d) As soon as it
reaches the destination, the vehicle is removed from the segment. The
travel time is recorded for statistical evaluation.

Note that all vehicles try to execute their route-plans independent of
the actual traffic conditions that they encounter along their way. In
heavily congested areas, vehicles often spill back across
intersections because they cannot enter their next respective
destination segments. In combination with the grid-oriented
characteristics of the CA model, this can result in complete
grid-locks of the simulation area, which cannot be resolved anymore
within the current rule-set. This artifact will be discussed in
paragraph \ref{results.gridlock}

\subsubsection*{Sources and Sinks}
Vehicles are inserted into and removed from streets at special sites
acting as sources and sinks. A source consists of $n_{source}$
consecutive sites (in each lane), which all have to be vacant before a
vehicle can be inserted. The location of a source (or sink) is usually
defined by intersections with the lower level street network that have
not been included into the simulation network. Thus, the sources serve
as aggregate feeding facilities.

The maximum throughput source is {\em one vehicle per time-step and
lane}, which is well above the rates seen in reality. In fact, these
values will never be reached, since the CA model usually restricts the
throughput of a lane to about 1600-1800 vehicles per hour. Remember
that the actual feeding frequency of a source, however, is determined
by the number of plans originating from it.

A sink consists of $v_{max}$ consecutive sites (on each lane) which
are scanned for vehicles that have reached their destination. This is
to make sure that every vehicle passing the sink spends at least one
time-step within this range. Since every vehicle is deleted without
delay, the sink can --- using the inverse argument as above --- absorb
all incoming traffic without generating a spill--back.

\section{Simulation Results}
\label{results}
\epsfxsize=10cm 
\eps{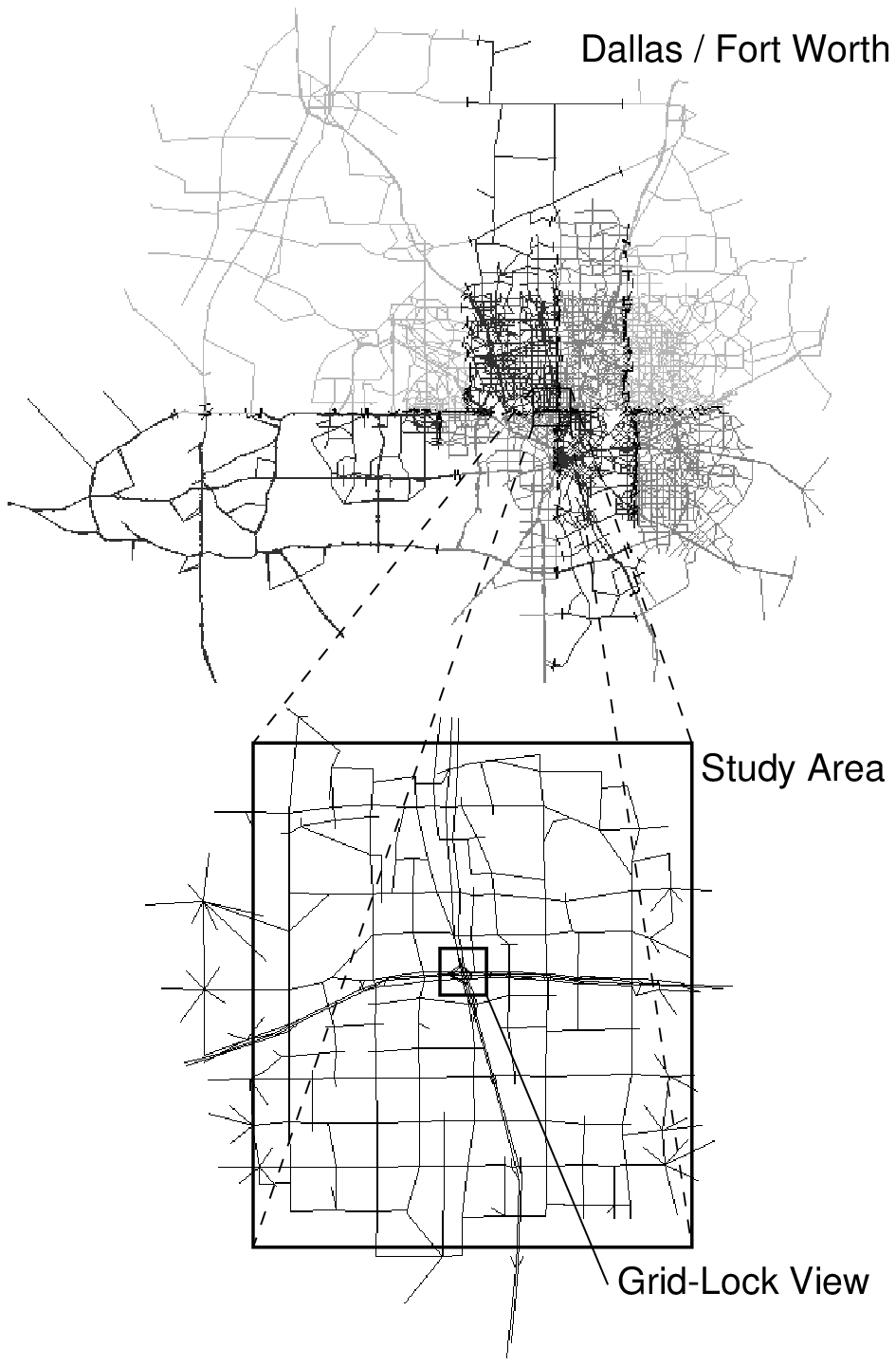}{Maps used in
simulation.  The different shades of gray in the Dallas/Fort Worth map
correspond to the mapping to different processors of the parallel
computer topology.}{fig_maps} 

All figures shown in this section are based upon data sets used in a case study of the
TRANSIMS \cite{transimsHTTP:1} project. We used two maps (see
\ref{fig_maps}): the complete Dallas/Fort Worth Area and a small
excerpt of the latter called {\em study area}. The study area maps
comprises all streets except small ones in residential areas and
similar areas. The large map further contains all minor and major
arterials for Dallas and all major arterials for Forth Worth.

The plan-sets which we had available contained only trip departure
times between 7\ a.m.\ and 10\ a.m.\ of which we selected those
between 7\ a.m.\ and 8\ a.m.\ as period of interest. We thus started
the simulation at 7\ a.m. and let it run at least until 8\ a.m. After
that, the simulation either terminated when (a) 99\% of all
route-plans had been executed, or (b) a grid-lock was detected. In
this context we assume the system to be grid-locked if the number of
vehicles in the system is constant for more than 600 time-steps. For
the CA model we used the deceleration probability of $p_d=0.3$ in all
simulations.

During the simulation we keep track of: the number of vehicles
inserted so far, the number of vehicles currently in the network, and
the number of vehicles that have reached their destination. Upon
arrival of each vehicle we store the estimated travel time (computed
by the router beforehand) and the actual travel time. These can be
compared to check the prediction quality of the router.

Except for the curves depicting the number of vehicles in the
study-area (figure~\ref{fig_plan11_vehicles}), we considered only vehicles
that arrived before time-step 1800. 
Also, all curves have been aggregated over 10 simulation-runs (using
different random seeds). They have been normalized by the respective number
of vehicles that have reached their destinations before time-step
1800. Moreover the area underneath each curve has been normalized to one.
\subsection{Levels of Fidelity}

It must be the principal goal of a traffic simulation within a cellular
automaton context to keep the interaction rule-set as small as
possible. This has two advantages: (a) The number of parameters will
be small, which reduces the probability of artifacts and makes the
model more credible, if it can be successfully validated using those
parameters. (b) A simple rule-set usually results in efficient coding,
which is essential, considering the before-mentioned necessity of
iteration and statistical averaging.

We examine different levels of fidelity by simulating the same
plan-set after activating different combinations of characteristics
of intersections and street segments which were described above.
We ran simulations for the combinations listed in
table \ref{tab_fidelity}.
\begin{tab}{|l|lllll|}{%
Parameters (bottom rows) defining the fidelity (right columns) of the simulation: In case of an
active speed-limit the maximum velocity $v_{max}$ is reduced from its original value of 5 to
a segment-specific value. In case of active traffic lights, the transfer at
intersections is decreased by introducing periodic red phases during which some
of the incoming segments are blocked. Reduced non-green phases
($q_r$) are discussed in paragraph \ref{reduced_non_green_phase_length}.
}
{tab_fidelity}
\hline 
short name & lf & sl & tl & hf & rl\\ 
long name & \underline{l}ow \underline{f}idelity & \underline{s}peed
\underline{l}imits & 
\underline{t}raffic \underline{l}ights & \underline{h}igh
\underline{f}idelity &
\underline{r}educed \underline{l}ights\\
\hline 
speed limit & no & yes & no & yes & yes\\ 
traffic lights & no & no & yes & yes & $q_r$\\ 
\hline
\end{tab}

\subsubsection*{Delay of arrival}
\epsfxsize=\xsize
\eps{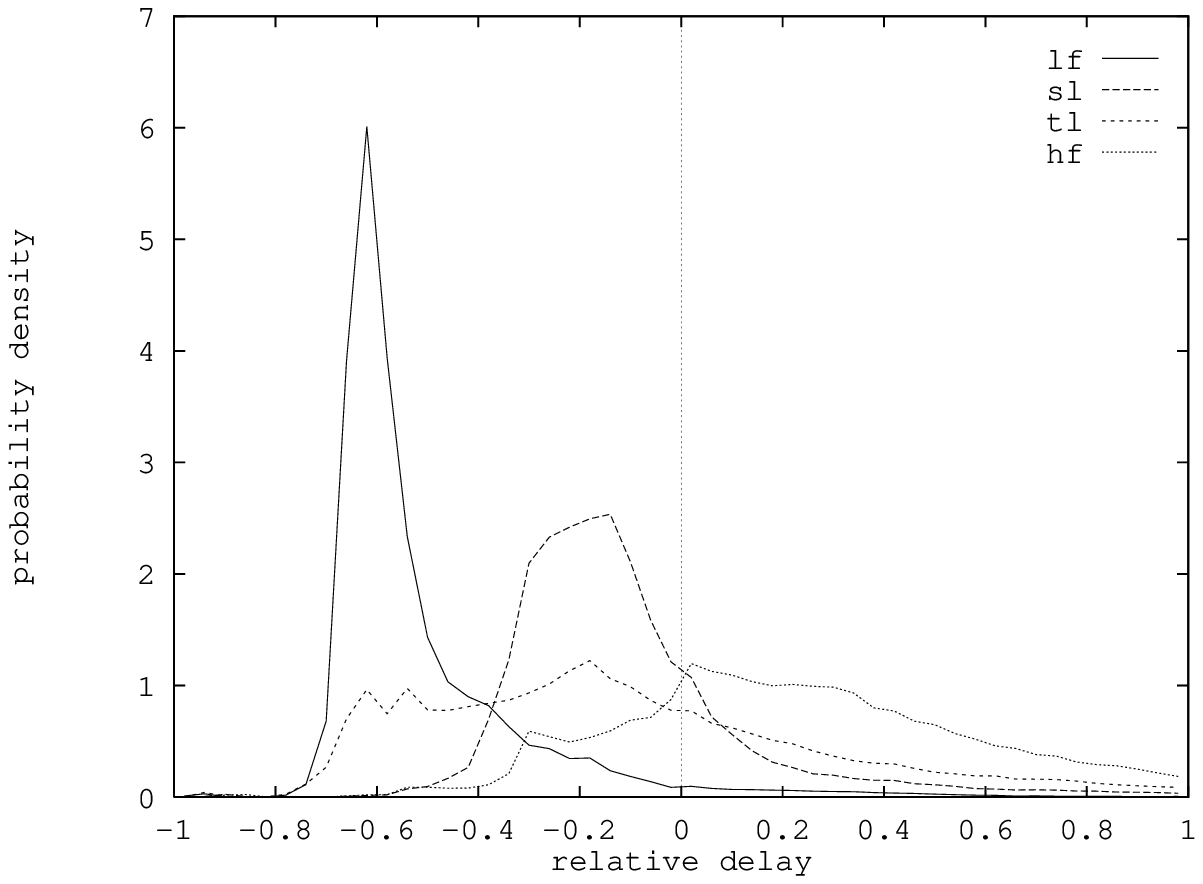}{Distribution of relative delays for plan-set 11}{plan_11_rel_delay}
\epsfxsize=\xsize
\eps{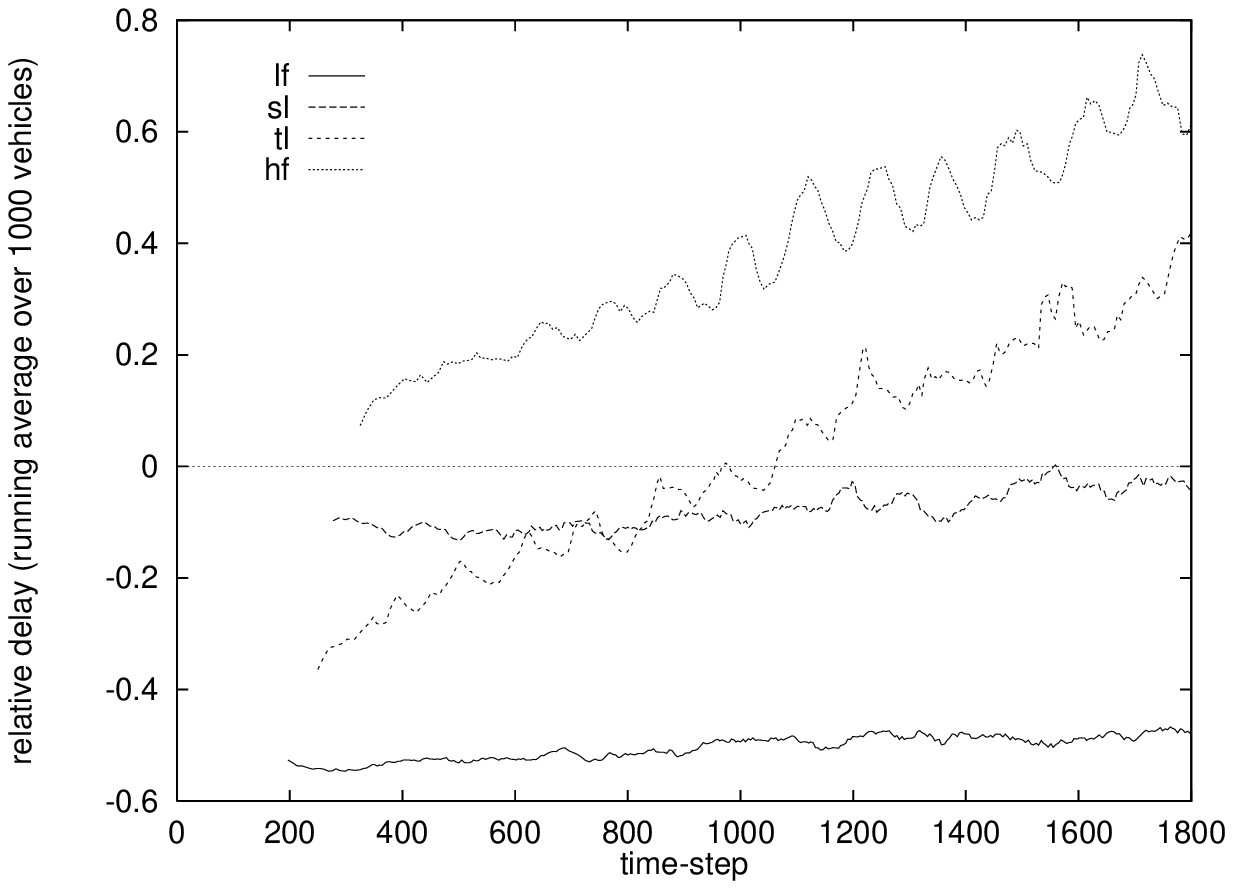}{Running average of relative delays for plan-set 11}{plan_11_rel_delay_ra}
Since upon arrival of each vehicle its actual travel time $t_{actual}$
is recorded, we could compute the distribution of relative delay $d$
\begin{displaymath}
d = \frac{t_{actual} - t_{scheduled}}{t_{scheduled}}
\end{displaymath}
with respect to the scheduled trip time $t_{scheduled}$ forecast by
the router. Note that negative values denote early arrivals. Figure\
\ref{plan_11_rel_delay} shows the results for plan-set 11 in all
fidelities.  It is obvious that in mode {\em lf} (due to the missing
speed-limit) route-plans are executed much too fast. There are hardly
any late-comers at all.  In modes {\em sl} and {\em tl} the peak is
already shifted towards zero delays but still biased. Mode {\em hf}
generates a distribution which peaks almost exactly at zero delay. The
average, however, is shifted towards positive delays. This can be
verified in figure~\ref{plan_11_rel_delay_ra}, which displays the
running average (over the latest 1000 vehicles) of the relative delay.
As can be seen, in the uncongested regime the travel times of the
microsimulation are in general shorter than what the planner had
expected.  Yet, in highly congested situations, travel times in the
microsimulation become longer than what the planner had expected.

We have to point out that it cannot be the goal of the dynamic
microsimulation to reproduce static results forecast by the
planner. These comparisons only serve as a consistency-check. It is to be
expected that results of the microsimulation will considerably
differ, e.g.\ in places where the implicit aggregation of the planner
smoothes over sudden peaks in traffic load.

\subsubsection*{Trip Duration}
\epsfxsize=\xsize
\eps{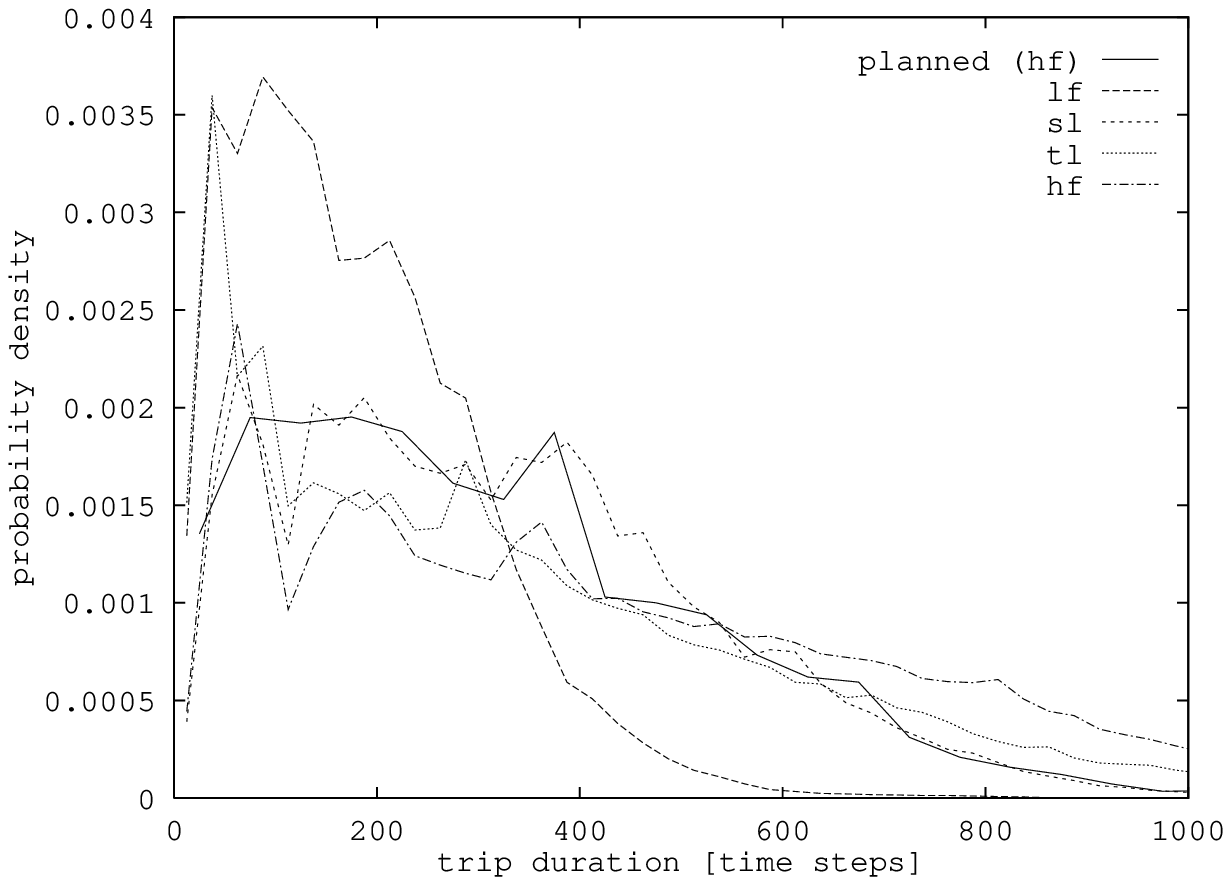}{Distribution of trip duration}{fig_trip_duration}

Figure\ \ref{fig_trip_duration} depicts the distribution of trip
durations for all fidelities. As expected mode {\em lf} has its peak
shifted far left towards small trip durations. Mode {\em tl} has a
significant peak at approximately 50 [sec], where it reaches the same
value as {\em lf}. This is due to the fact that the probability of
encountering a traffic light is a very small for short trips,
rendering modes {\em lf} and {\em tl} equivalent in that regime. Modes
{\em tl} and {\em hf} have a very slow descent towards large trip
durations due to grid-locks. The curve of planned trip duration is
best matched by mode {\em sl}.

\subsection{Reproducibility and Grid-Locks}
\label{results.gridlock}
\epsfxsize=\xsize
\eps{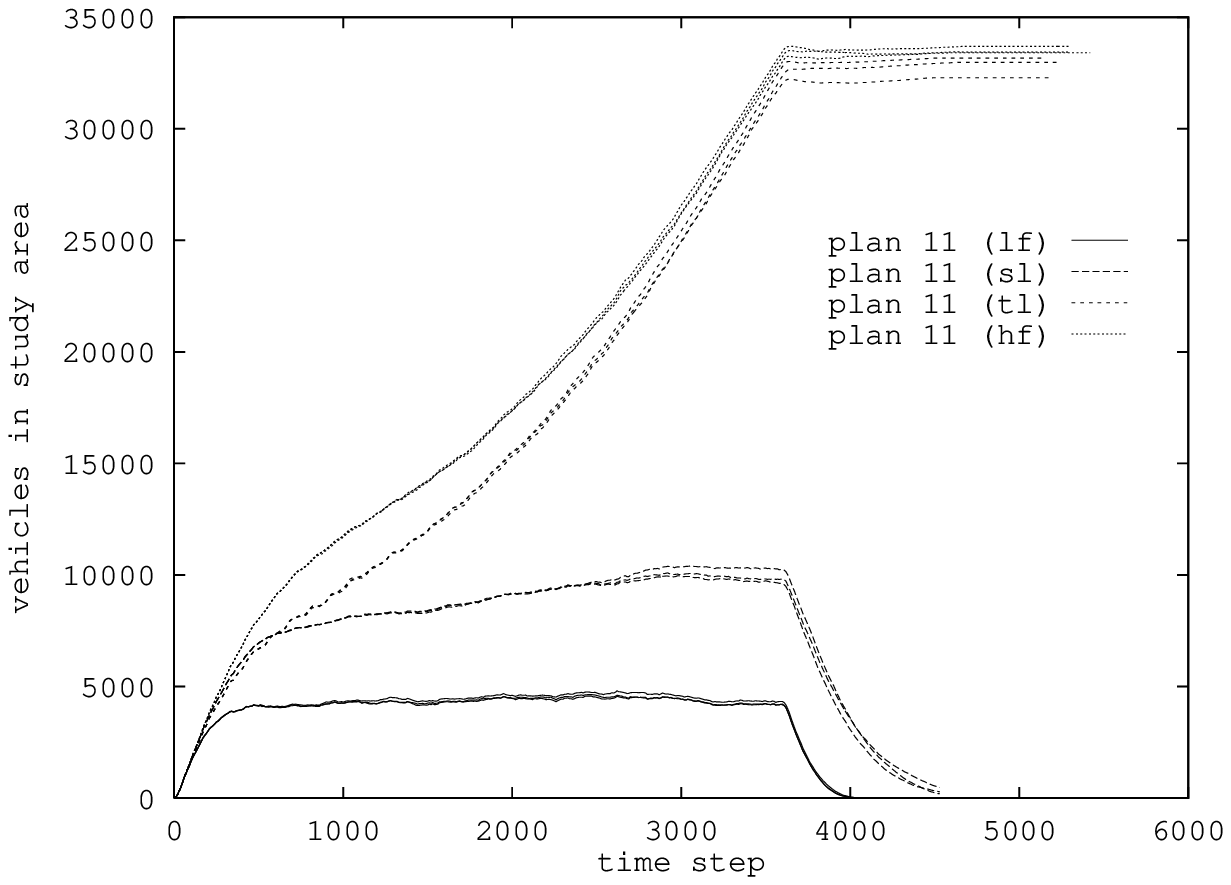}{Vehicles in study area for plan-set
11}{fig_plan11_vehicles}

Since the CA model contains a stochastic element, we receive a unique
evolution of the simulation for each seed of the random number
generator.  In a sub-critical system the network is able to transport
all vehicles (albeit with delay) so that all runs will look similar on
a macroscopic level.  In a system with a network throughput incapable
of handling the loading, the system will most likely grid-lock (see
below). Between the two extremes we find a regime in which the
specific configuration may either block or not block.  Figure\
\ref{fig_plan11_vehicles} depicts the number of vehicles which are in
the study-area at a given time-step. Fidelities {\em lf} and {\em sl}
belong to the sub-critical regime. Both curves reach a plateau (at
4500 vehicles after time-step 500 for {\em lf} and at 10000 vehicles
after time 2500 for {\em sl}) after an initial loading phase
representing an equilibrium between the insertion and deletion rates
of vehicles. After the loading phase all remaining vehicles are
discharged within 400 time-steps for {\em lf} and within 900
time-steps for {\em sl}.

Modes {\em tl} and {\em hf} belong to the super-critical regime. They
never reach an equilibrium between insertion and deletion during the
loading phase; and the plateau after the loading of the network is due
to grid-lock.

See Ref.~\cite{IVHS-testbed} for a similar (albeit much smaller scale)
investigation on the relation between network loading and network
throughput. 

\subsubsection*{Grid-locks}
\epsfxsize=\xsize
\eps{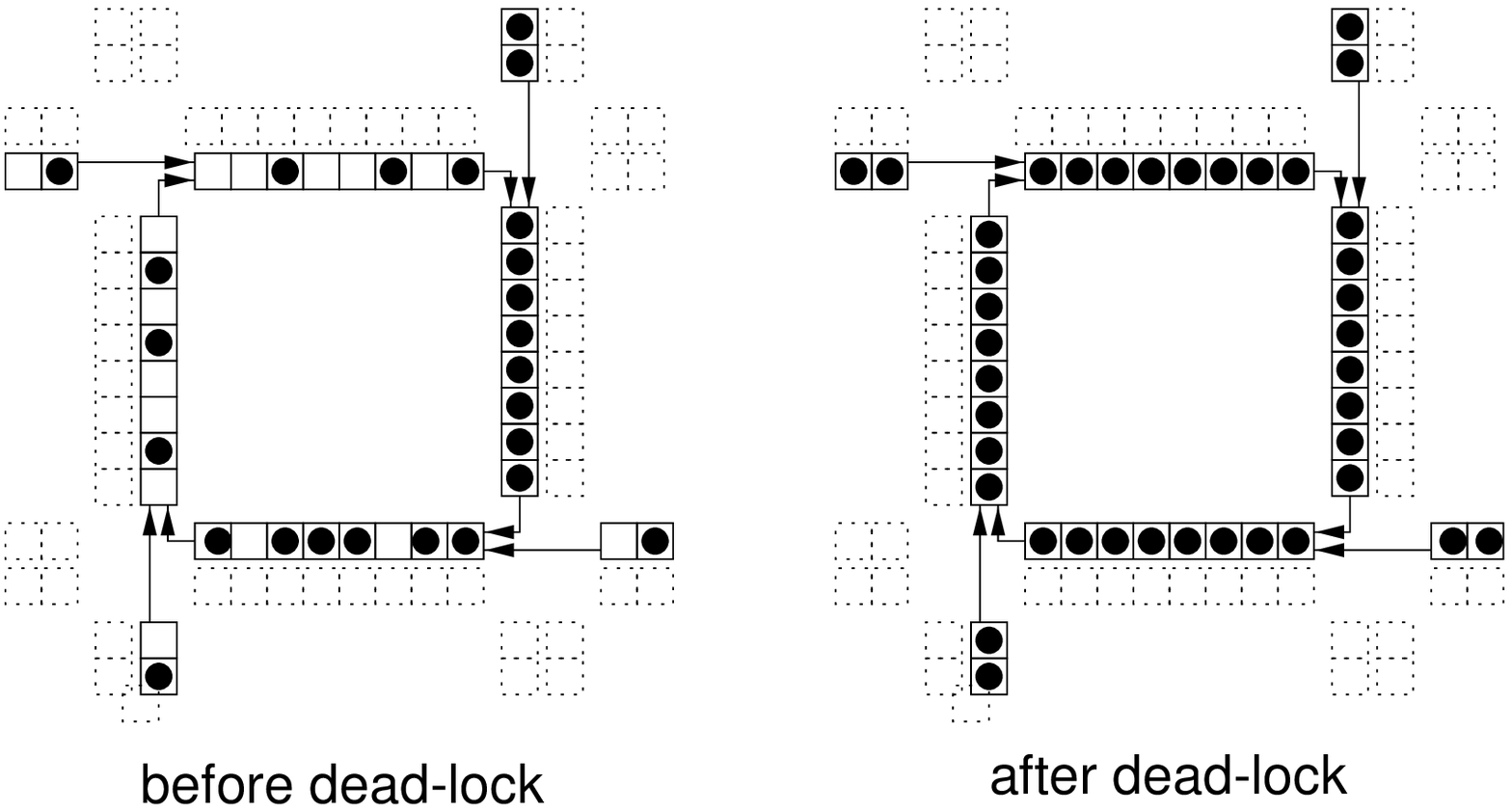}{Grid-lock}{fig_gridlock}
\epsfxsize=\bitxsize
\eps{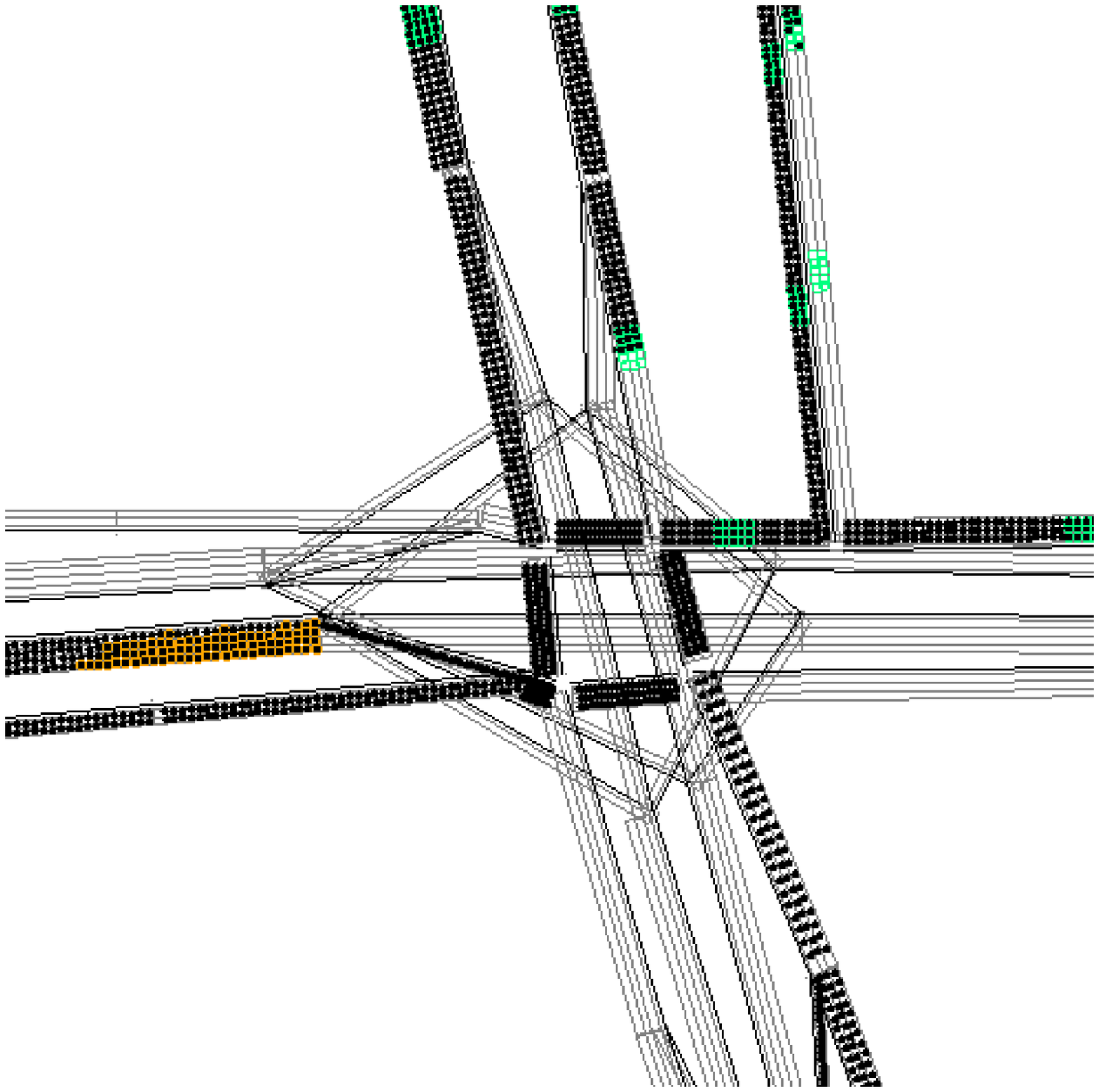}{Grid-lock in study-area (plan 11,
mode hf)}{fig_plan11_hf_gridlock}
A grid-lock situation can be determined by a horizontal line after
time-step 3600 (e.g. the end of vehicle insertion). This is caused by
closed loops in the traffic network in which all sites are
occupied. Figure \ref{fig_gridlock} depicts a simplified
intersection. In the left half traffic is already dense, though not
grid-locked. Due to high demand and the red-phases at the
intersection, the segments of the loop are no longer cleared. In the
right half the whole loop is blocked: the first vehicle in each lane
is forced to make a right turn into another lane which is also
blocked. This phenomenon (in its strict form) cannot be seen in
real-world traffic, because drivers move out of the lanes and pass on
the on-coming lane or they abandon their current route and choose a
detour. Figure\ \ref{fig_plan11_hf_gridlock} shows a screen-shot of a
simulation run with plan-set 11 in mode {\em hf}. Vehicles are
represented by dark dots, lane boundaries by grey lines. Right at the
center, there is a small grid-locked loop blocking traffic from all
incoming directions.

\subsection{Reduced Non-Green Phase Length}
\label{reduced_non_green_phase_length}
\epsfxsize=\xsize
\eps{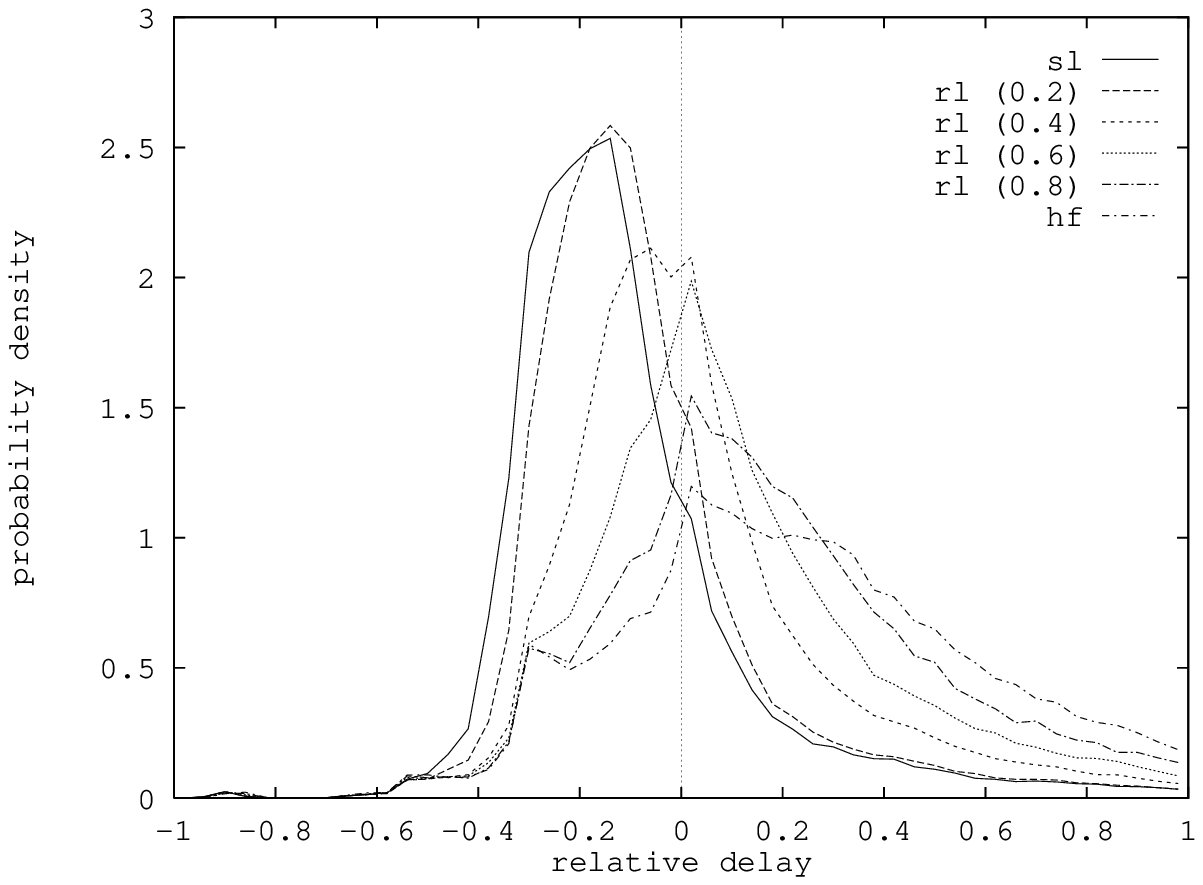}{Distribution of relative delay of plan-set 11 (different
$q_r$)}{fig_plan11_rl_delay}
\epsfxsize=\xsize
\eps{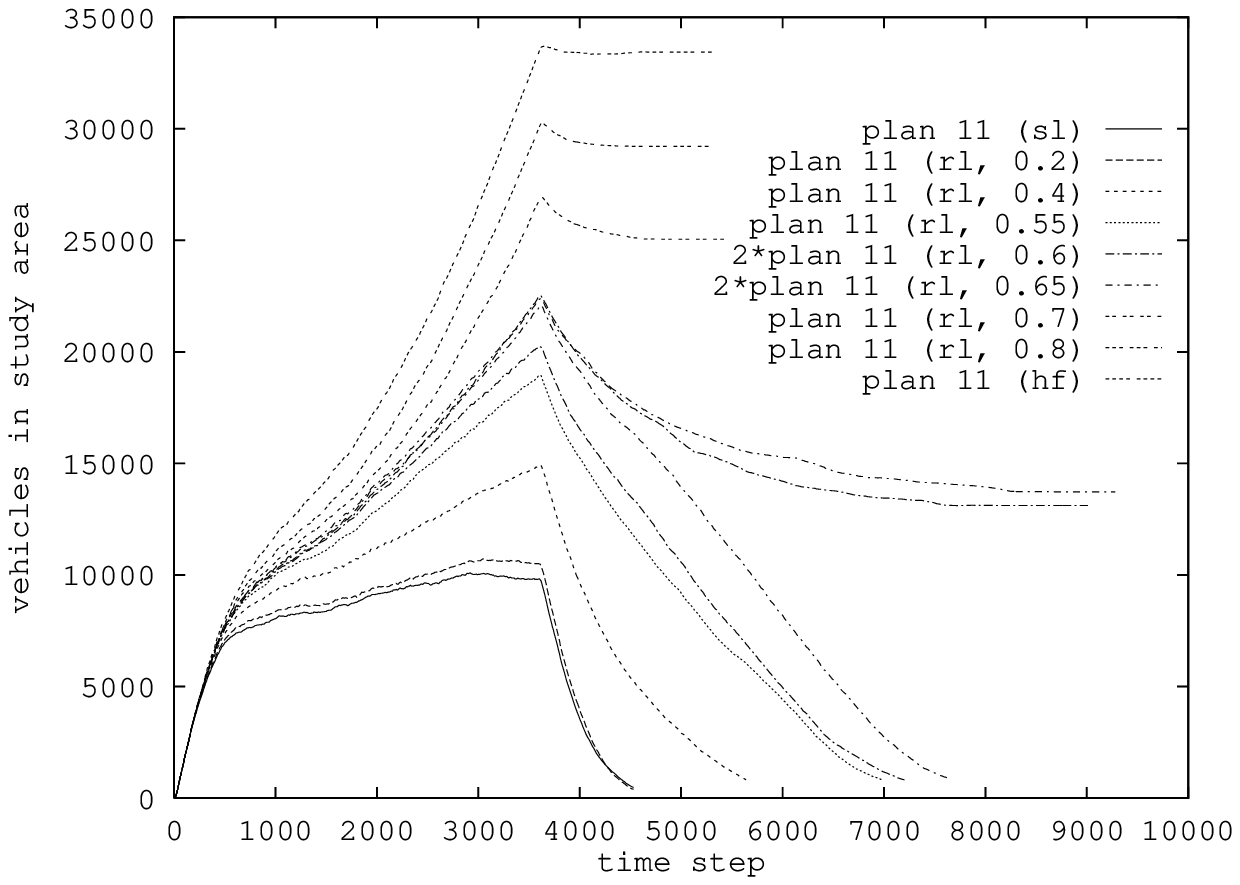}{Vehicles in study area (different
$q_r$)}{fig_plan11_rl_vehicles}
\epsfxsize=\xsize
\eps{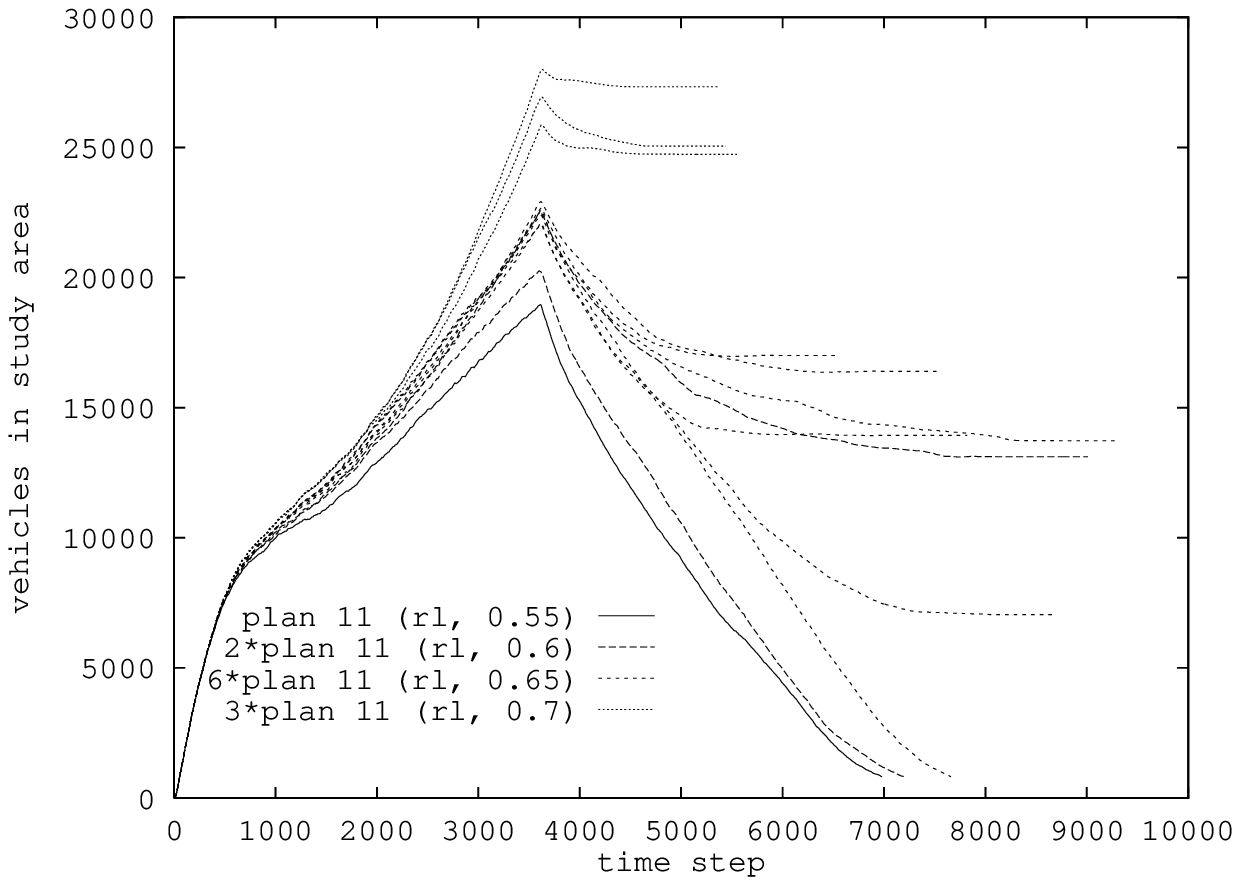}{Vehicles in study area
($q_r=0.55, 0.6, 0.65, 0.7$)}{fig_plan11_rl_vehicles_gridlock}

Above we have seen cases of sub-critical and super-critical loading of
the network. Although the respective models are defined by different
types of rule-sets, they can be regarded as two specific cases of a
more general rule-set (called {\em rl}) in which in the effective
red-phase $T_{r,eff}$ is computed by multiplying the original
phase-length $T_r$ by a certain factor $q_r\in [0\ldots
1]$. Consequently, $q_r=0$ represents mode {\em sl} while $q_r=1$
represents mode {\em hf}.

We have made several runs for different values of $q_r$ between 0 and
1. Both figures \ref{fig_plan11_rl_delay} and
\ref{fig_plan11_rl_vehicles} show a smooth transition between modes
{\em sl} and {\em hf} (for $0.6$ and $0.65$ blocking and non-blocking
representatives were chosen) as far as delay and trip duration are
concerned. Values below $0.6$ show a secure sub-critical (no
grid-locks), and values above $0.65$ a secure super-critical behavior. For
$0.6$ and $0.65$ the system has a certain chance of reaching a
grid-lock (1 out of 10 runs for $q_r=0.6$ and 5 out of 10 for
$q_r=0.65$), which can better be seen in figure\
\ref{fig_plan11_rl_vehicles_gridlock}.

A similar effect was reported for simple 2-dimensional grid
models~\cite{Cuesta:etc,Freund:Poeschel,nagatani93b}, except that in
these studies the overall density was changed instead of the
efficiency of the network components.  Intuitively, the grid-lock
effect seems to be the same.  Yet, further investigations will be
necessary to understand in how far simple models on a 2-dimensional
grid can indeed offer insight for real-world city traffic, which is
admittedly happening in 2-dimensional space but is composed of traffic
on 1-dimensional links.  It is for example unclear if dynamical
critical exponents are the same.

\section{Performance and Implementation}
\label{performance}
\label{implementation}
\begin{tab}{|l|cccc||c|}{Performance}{tab_mosps}
\hline
Map            & \multicolumn{4}{c||}{Study Area} & Dallas / Fort Worth \\
Architecture   & \multicolumn{4}{c||}{SUN Sparc 20} & 8 SUN Sparc 20 \\
Resolution     &   lf    &   sl    &   tl    &   hf    & hf \\
\hline
MOSPS (average)&  0.015  &  0.032  &  0.077  &  0.092  & 0.095\\
MOSPS (max)    &  0.022  &  0.045  &  0.125  &  0.153  & 0.156\\
MUPS (average) &  0.704  &  0.689  &  0.788  &  0.647  & 4.880\\
MUPS (max)     &  2.020  &  1.989  &  1.953  &  1.908  & 8.950\\
\hline
\end{tab}
We measure performance in units of the so-called real-time-ratio
(RTR). An RTR of 1.0 means that one simulation second (corresponding
to one simulation time-step) takes one wall-clock second to
execute. Larger values denote faster simulations. Other interesting
measurement units are the number of vehicles (MOSPS = million object
seconds per second) and the number of sites (MUPS = million updates
per second) that can be updated in one wall-clock second. Table\
\ref{tab_mosps} contains an overview of all performance
benchmarks. Note that the MOSPS values are density-dependent which
results in larger numbers for grid-locking modes {\em tl} and {\em hf}
since the network contains up to three times as many vehicles.

\subsection{Study Area (1 CPN)}
\epsfxsize=\xsize
\eps{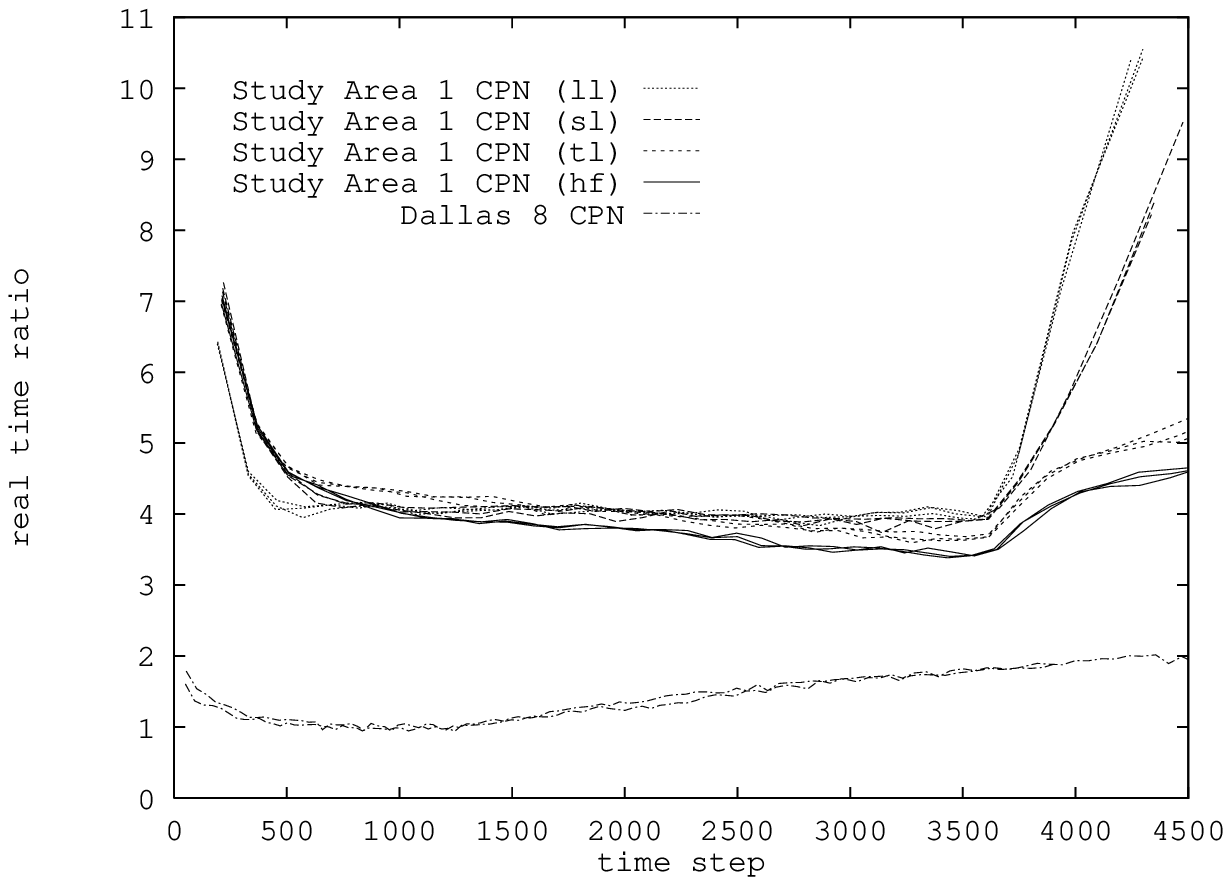}{Performance for study area map}{fig_study_area_perf}
The map excerpt representing the study-area comprises of 374
intersections (93 of which have traffic lights) and 528 segments (142
of which are one-way). The total length of the roadway summed over all
lanes is 1154 [kilometer] or 153,022 sites.  All runs were done on
single Sparc Stations 5 and 20. Figure\ \ref{fig_study_area_perf}
shows the performances for all fidelities during the simulation. The
RTR drops rapidly until approximately 5000 vehicles are in the system
and more slowly until the end of the vehicle insertion phase is
reached. As expected, the high fidelity mode is the slowest. The
difference between the modes is minimal and the RTR does not drop far
below four during the main portion of the simulation.

\subsection{Dallas/Fort Worth (8 CPN)}
\epsfxsize=\bitxsize
\eps{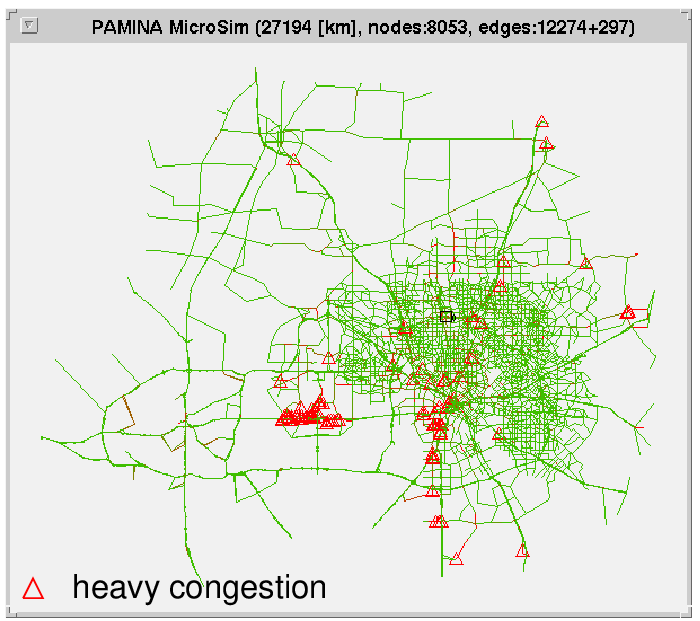}{Screen-shot of congestion in Dallas/Fort
Worth area after 9 minutes}{fig_dallas_9min}
The map representing the whole Dallas/Fort Worth area comprises of
8053 intersections (93 of which have traffic lights = those in the
study area) and 12,274 segments (4772 of which are one-way and 366 of
which were split across CPN boundaries). The total length of the
roadway summed over all lanes is 27,194~kilometers or 3,608,781 sites.

The whole Dallas/ Fort Worth Area was simulated on a SUN Sparc
workstation cluster using 8 machines coupled via optical LAN (see
figure\ \ref{fig_maps}). The RTR was approximately 1.0 between
time-steps 500 and 1200 when the number of vehicles reached its
maximum at 128,000 vehicles. Unlike before, the route set contained all
routes through the region. Disk space limitations forced us to shorten
the insertion period to 20 minutes: Using the current TRANSIMS plans
format, plans for one hour would have needed close to one GByte of
disk space.  The results presented here shall thus only serve as a
proof of feasibility. For the time being, a more detailed evaluation is not reasonable,
since the underlying network was not homogeneous: outside Dallas only
major arterials were used to hold the traffic which is usually spread
over the topmost two or three street hierarchies. We experienced
extreme clogging of those arterials. A good estimate for the
computational speed of the overall detailed network can be obtained if
the overall length (in lane kilometers ) of the network is
known. Usually the factor between a high-fidelity and a arterial
network is about 3 to 5 which would still result in a RTR of about 0.2-0.3
for this type of architecture.

Figure\ \ref{fig_dallas_9min} depicts the network after 9
minutes of loading. The small triangles denote heavy congestion. Note
the heavy congestion along the southbound and westbound axes.

\subsubsection*{Parallelization}
We use a geometrical domain decomposition approach to distribute
the street network among the CPN. A recursive orthogonal bisection is
used to determine the sub-networks which are connected through
boundaries located in the middle of street-segments. Before each
time-step boundary information consisting of vehicle state information
is passed to neighboring CPN with message passing routines
(PVM\cite{PVMHTTP:1}). The time-step itself is independently performed in parallel
on all machines. For a more detailed description of parallelization
scheme see \cite{RiWa:1,RiWaGa:1}.

\section{Summary, Conclusion and Future Work}

We have presented a simple microsimulation model for city
traffic. The goal was to find a simple extension to the original
traffic CA which can handle city-like intersections. In
its ``high fidelity'' mode, the proposed intersection model reaches
satisfactory results while maintaining a high computational speed. 

We have pointed out the danger of grid-locks which are due to the
granular structure in combination with static route-plans. It will be
necessary to include on-line re-routing to reduce the risk of
grid-locks. Also, the impact of travellers' information systems can
only be examined, if route-plans are no longer static for the duration of
one simulation run.

We showed that the a simple parameter like
$q_r$ can be used to fine-tune the performance of the intersection
throughput. For $q_r\simeq 0.6$ the curves for relative delay were
optimal while the behavior with respect to grid-locking was improved
considerably.

In future, we will repeat the runs with more detailed maps: on the one
hand, it is necessary to determine what fidelity the enclosing network
(that is Dallas/Fort Worth {\em without} the study area) will have to
have to suppress the congestion outside the study area. On the other
hand, the map of the study area which was used for all simulations in
this paper did still not include {\em all} street levels. The system
actually was missing some of its road capacity located in residential
areas. The sources and sinks acted as devices which aggregated (or
disaggregated respectively) the traffic flow close to insertion and
deletion points. As a result, the existing streets are not loaded as
homogeneously as they would be, if all access points (sources or
sinks) only represented their immediate vicinities.

Running the simulation multiple times reveals information about the
robustness of traffic flow along specific routes. It will be
interesting to classify street segments by the fluctuation of traffic
flow data collected over several runs. Areas of high fluctuations may
be good candidates to install intelligent traffic control mechanism to
prevent congestion using on-line traffic counts. Another interesting
aspect would be the examination of damage spreading in a complex
traffic system: reducing the efficiency or blocking an intersection
for a certain period will cause traffic jams moving away from the
origin of the disturbance and possibly changing the fluctuation structure
in another region of the network.

\section*{Acknowledgements}
We thank A.~Bachem, R.~Schrader and C.~Barrett for supporting MR's and
KN's work as part of the traffic simulation efforts in Cologne
(``Forschungsverbund Verkehr und Umwelteinwirkungen NRW''
\cite{nrwfvuHTTP:1}) and Los Alamos (TRANSIMS). We also thank them,
plus P.~Wagner, Ch.~Gawron, S.~Krau{\ss}, S.~Rasmussen, and
M.~Schreckenberg for help and discussions.  We thank the TRANSIMS
research (M.~Marathe and others) and software (M.~Stein, P.~Stretz,
D.~Roberts, B.~Bush, K.~Bergbigler, and others) teams for discussions
and the processing of the data which served as input for the micro
simulation.  North Central Texas Council of Government (NCTCOG)
provided road network and origin-destination data on which the
simulations were based.  Use of computer facilities at the Center of
Parallel Computing (Cologne) and computing time on the workstation
cluster at TSA-DO/SA is gratefully acknowledged. We further thank all
people in charge of maintaining the above mentioned cluster.

The work of MR was supported in part by the ``Graduiertenkolleg
Scientific Computing K\"oln''.



\end{document}